\documentstyle[amstex,amssymb,preprint,prd,aps]{revtex}
\begin{document}
\tightenlines
\draft
\title{Scaling Solutions and Reconstruction of Scalar Field Potentials}
\author{Claudio Rubano$^{1}$\medskip\ and John D. Barrow$^{2}$}
\address{ $^{1}$ Dept. of Phys. Sciences - Univ. Federico II
 and INFN Sez. di Napoli
 \\ Complesso Universitario di Monte S.
Angelo, \\ via Cintia, Ed. G, I-80126 Napoli - Italy \\
\smallskip
$^{2}$ DAMTP, Centre for Mathematical Sciences,  Cambridge University,\\
WilberforceRd., Cambridge CB3 0WA, U.K.}
\smallskip
\date{\today}
\maketitle

\begin{abstract}
Starting from the hypothesis of scaling solutions, the general exact form
of the scalar field potential is found. In the case of two fluids, it turns
out to be a negative power of hyperbolic sine. In the case of three fluids
the analytic form is not found, but is obtained by quadratures.

\end{abstract}

\subsection*{Introduction}

\noindent\ \ \ In cosmological theories containing scalar fields with a
self-interaction potential $V(\varphi),$ it is sometimes possible to
reconstruct the required scalar field potential for a simple cosmological
solution. In the context of inflationary theory, this approach was used by
various authors \cite{ellis}, \cite{barr1}, \cite{mang}, \cite{deritis},
\cite{BP},\cite{jantzen1} \cite{jantzen2} \cite{jantzen3}
\cite{capozziello1}, \cite{capozziello2}, and was primarily interested in
the behaviour of solutions containing only scalar fields undergoing
inflation. The more recent invocation of a scalar field as a dark matter
source responsible for accelerating the universe today, under the pseudonym
of ``quintessence'', \cite{ostriker}, \cite{caldwell}, \cite {zlatev},
\cite{steinhardt}, \cite{barr2}, is mathematically almost identical, but
places different requirements on the solutions. In particular, it is of
interest to find solutions which contain both perfect fluids and scalar
fields. In this new context, interesting applications of the reconstruction
approach were made by Chiba and Nakamura \cite{chiba} and Saini et al.
\cite{saini}.

\ \ \ \ \ \ In this paper, we seek exact cosmological solutions for a
universe containing a perfect fluid and a scalar field. We start from the
assumption that the energy density of the scalar field scales as an exact
power of the scale factor: $\rho_{\varphi}=K\,a^{-n}$ , which is equivalent
to imposing an equation of state linking the pressure and density, of the
form $p_{\varphi }=w\rho_{\varphi}$ with constant $w=n/3-1$. For a flat
Friedmann universe, it is then possible to find an explicit exact form of
the potential in terms of $n$, $H_{0}$ , and $\Omega_{m0}$. The form of the
resulting general solution has instructive features which will be discussed
below.

\subsection*{Derivation of the Potential}

Consider a cosmological fluid with two non-interacting components:
perfect-fluid matter and a scalar field $\varphi$ with potential $V(\varphi)$%
. \ In the flat universe case, we have the equations
\begin{gather}
3H^{2}={\cal G(\rho}_{m}+{\cal \rho}_{\varphi}) \\
\ddot{\varphi}+3H\dot{\varphi}+V^{\prime}(\varphi)=0 \\
{\cal \rho}_{\varphi}=\frac{1}{2}\dot{\varphi}^{2}+V(\varphi) \\
{\cal \rho}_{m}=Da^{-m}\ ,
\end{gather}
where $H=\dot{a}/a$, where $a(t)$ is the expansion scale factor is the
Hubble expansion rate, and overdot denotes differentiation with respect to
the comoving proper time $t$\/; ${\cal G}=8\pi G/c^{2}$ and $V^{\prime
}=dV/d\varphi$. The constant $m$ depends on the type of perfect fluid
present.

We normalize the present value of the scale factor to $a_{0}=1$, without
loss of generality, and for brevity denote the present matter-density
parameter, $\Omega_{m0},$ by $\Omega_{0}$, and define
\begin{equation}
D=3H_{0}^{2}\Omega_{0}/{\cal G\ }.
\end{equation}

If we make the assumption that
\begin{equation}
{\cal \rho}_{\varphi}=Ka^{-n},
\end{equation}
with $n<m$\/, so the scalar field can dominate at late times, and define
\begin{equation}
K=3H_{0}^{2}(1-\Omega_{0})/{\cal G\ },
\end{equation}
then from Eqs. (2), (3) and (6) we obtain
\begin{equation}
\dot{\varphi}^{2}=\frac{Kn}{3}a^{-n}.
\end{equation}
Since
\begin{equation}
\frac{d\varphi}{dt}=Ha\frac{d\varphi}{da}\ ,
\end{equation}
we have
\begin{equation}
\left( \frac{d\varphi}{da}\right) ^{2}=\frac{K}{3H_{0}^{2}}\frac{\sqrt{n}}{%
\Omega_{0}a^{n-m+2}+(1-\Omega_{0})a^{2}}\ ,
\end{equation}
which gives
\begin{multline}
\varphi(a)=\int\sqrt{\frac{1-\Omega_{0}}{{\cal G}}}\frac{\sqrt{n}\,da}{\sqrt{%
\Omega_{0}a^{n-m+2}+(1-\Omega_{0})a^{2}}} \\
=\frac{2\sqrt{n}}{\sqrt{{\cal G}}(m-n)}{\rm arc}\sinh\left( \sqrt {\frac{%
1-\Omega_{0}}{\Omega_{0}}}a^{\frac{m-n}{2}}\right) +\varphi_{in}\ .
\end{multline}

Returning to the potential, we get from Eqs. (3) and (9)
\begin{equation}
V=Ka^{-n}-\frac{1}{2}\dot{\varphi}^{2}=\frac{3H_{0}^{2}}{{\cal G}}\left(
1-\Omega_{0}\right) \left( 1-\frac{n}{6}\right) a^{-n}\ .
\end{equation}

The scale factor can be easily eliminated, giving eventually
\begin{multline}
V(\varphi)=\frac{3H_{0}^{2}}{{\cal G}}\left( 1-\Omega_{0}\right) \left( 1-%
\frac{n}{6}\right) \left( \frac{1-\Omega_{0}}{\Omega_{0}}\right) ^{\frac{n}{%
m-n}} \\
{\times}\left( \sinh\left( \sqrt{{\cal G}}\frac{m-n}{\sqrt{n}}%
(\varphi-\varphi_{in})\right) \right) ^{-\frac{2n}{m-n}}.   \label{pot}
\end{multline}

This expression, in the case of dust ($m=3$), coincides with the one
presented by  \cite{urena} (which is in turn a particular case of the
treatment of \cite{chimento})as well as with the one by \cite{sahni}. A
 general discussion about exact solutions for Friedmann
equations, which includes ours as a particular case, can be found also in
\cite{jantzen2}, \cite{jantzen3}, \cite{jantzen1}. In \cite{urena} and
\cite{chimento}, one can find explicit solutions for $a(t)$ and
$\varphi(t)$, as well as an extensive discussion. We remark only that this
form of potential is good for a tracker solution \cite{zlatev},
\cite{steinhardt}. Indeed, straightforward computation of the function
$\Gamma = V^{\prime\prime}V/(V^{\prime})^2 $, introduced and discussed in
these papers, gives
\begin{equation}
\Gamma = 1+ \frac{m-n}{2 n} \left ( {\rm sech\/} \left ( \frac{m-n}{\sqrt n}
\varphi \right ) \right)^2 > 1
\end{equation}
as required by the tracking condition.

Our derivation differs from \cite{urena}, \cite{chimento} and \cite{sahni}
because it is  simpler and is generalised to include all perfect fluid
equations of state (other than the $n=6$ case which would correspond to a
pure scalar field with no potential). Moreover, it proves that this form of
$V(\varphi)$ is the {\em unique} solution, if condition (6) is imposed.

\subsection*{Discussion}

\ The first interesting feature of Eq. (\ref{pot}) is that the slope (as
well as the amplitude) of $V(\varphi)$ depends on $n$. This means that it is
impossible to obtain a scaling solution, with the same potential slope, when
passing from a radiation-dominated $(m=4)$ epoch to a matter-dominated $(m=3)
$ epoch. Even if we assume that $n$ changes in such a way that the slope
remains constant, the coefficient $(1-n/6)$ changes. Moreover, the effective
equation of state of the scalar field also changes and there is no physical
mechanism for it to be influenced in this way by the dominating type of
matter. Although this situation seems to be unphysical, it is merely an
artifact of having sought a solution containing a single perfect fluid. The
full solution must be found by including dust, radiation and scalar field
from the outset, not by joining the radiation + scalar solution to the dust
+ scalar solution. If this is done for the case of dust + radiation + scalar
field, with the same arguments as before, \ it is easy to derive
\begin{align}
\varphi & =\int\sqrt{\frac{\Omega_{\varphi}}{{\cal G}}}\frac{\sqrt{n}\,da}{%
\sqrt{\Omega_{r}a^{n-2}+\Omega_{d}a^{n-1}+\Omega_{\varphi}a^{2}}}%
+\varphi_{in} \\
V & =\frac{3H_{0}^{2}}{{\cal G}}\Omega_{\varphi}\left( 1-\frac{n}{6}\right)
a^{-n}\ ,
\end{align}
where $\Omega_{r}$, $\Omega_{d}$ and $\Omega_{\varphi}$ are the present
values of the radiation, dust and scalar field density parameters,
respectively, so $\Omega_{r}+\Omega_{d}+\Omega_{\varphi}=1$. These equations
give a parametric representation of $V(\varphi)$, which cannot be solved
analytically with simple functions, but is otherwise perfectly well defined
and interpolates between the dust and radiation solutions of Eq. (\ref{pot}%
). In our universe today, $\Omega_{r}<<\Omega_{d}$, but at early times the
radiation term $\Omega_{r}a^{n-2}$ dominates the dust and cannot be dropped.

A numerical example illustrates the situation: let us set $n=1$, $%
\Omega_{\varphi}=0.7$, $\Omega_{r}=0.0001$, $\Omega_{d}=0.3-\Omega_{r}$, $%
3H_{0}^{2}=1$, ${\cal G}=1$. Fig. 1 shows that on the first part the
``true'' potential fits well with Eq. (\ref{pot}) and $m=4$, while Fig. 2
shows that in the late regime the fit should be done with $m=3$.

The simple and attractive form of Eq. (\ref{pot}) is lost, although it might
be recovered by a suitable choice of the exponent: via some weighted mean
with of $\Omega_{r}$ and $\Omega_{d}$. But there is no way of doing this
other than a fit of the numerical values. Moreover, there is no reason why
the exponent should depend on the particular values of $\Omega_{r}$ and $%
\Omega_{d}$. Also, since in this case the scaling feature of the solution is
only approximate, the tracker behavior could be affected.

\medskip\ \ \ \ \ \ Simple numerical evaluations (with the parameters within
the allowed range) show that the ``true'' potential is very well
approximated by
\begin{equation}
V=\frac{\alpha}{\varphi^{\beta}}-\gamma\ ,
\end{equation}
with $\alpha$, $\beta$, $\gamma$ {\em positive} constants depending on the
parameters; in particular, $\beta$ $\approx$ $2n/(m-n)$. The negative
additive term, $\gamma$, is an artifact of the approximation over a finite
range of $\varphi$. This result, already found by \cite{urena} for the dust
case, shows that the inverse-power potential is effectively equivalent to
Eq. (\ref{pot}), so that it is possible to apply to this situation all the
known results about tracker solutions. On this point, it is interesting to
note that this approximation is very good over the whole range of $a$ from
zero to the present-day value ($a_{0}=1$ according to our normalization).
The asymptotic exponential behavior of the hyperbolic potential is therefore
important only in the very far future and does not affect the dominance of
scalar field now or its behaviour in the recent past.

\medskip\ \ \ \ \ As a final remark, we should stress that our arguments are
based on the arbitrary assumption of Eq. (6). We have shown elsewhere \cite
{rubano} that other forms of exponential potential are perfectly able to
reproduce observational data, but of course in these cases $w$ is not
constant. It is interesting also to note that, in one of the cases treated
in that paper, $w$ is almost perfectly constant, and yet the form of the
potential is substantially different from that of Eq. (\ref{pot}).

\bigskip

\bigskip

{\bf Figure Captions}

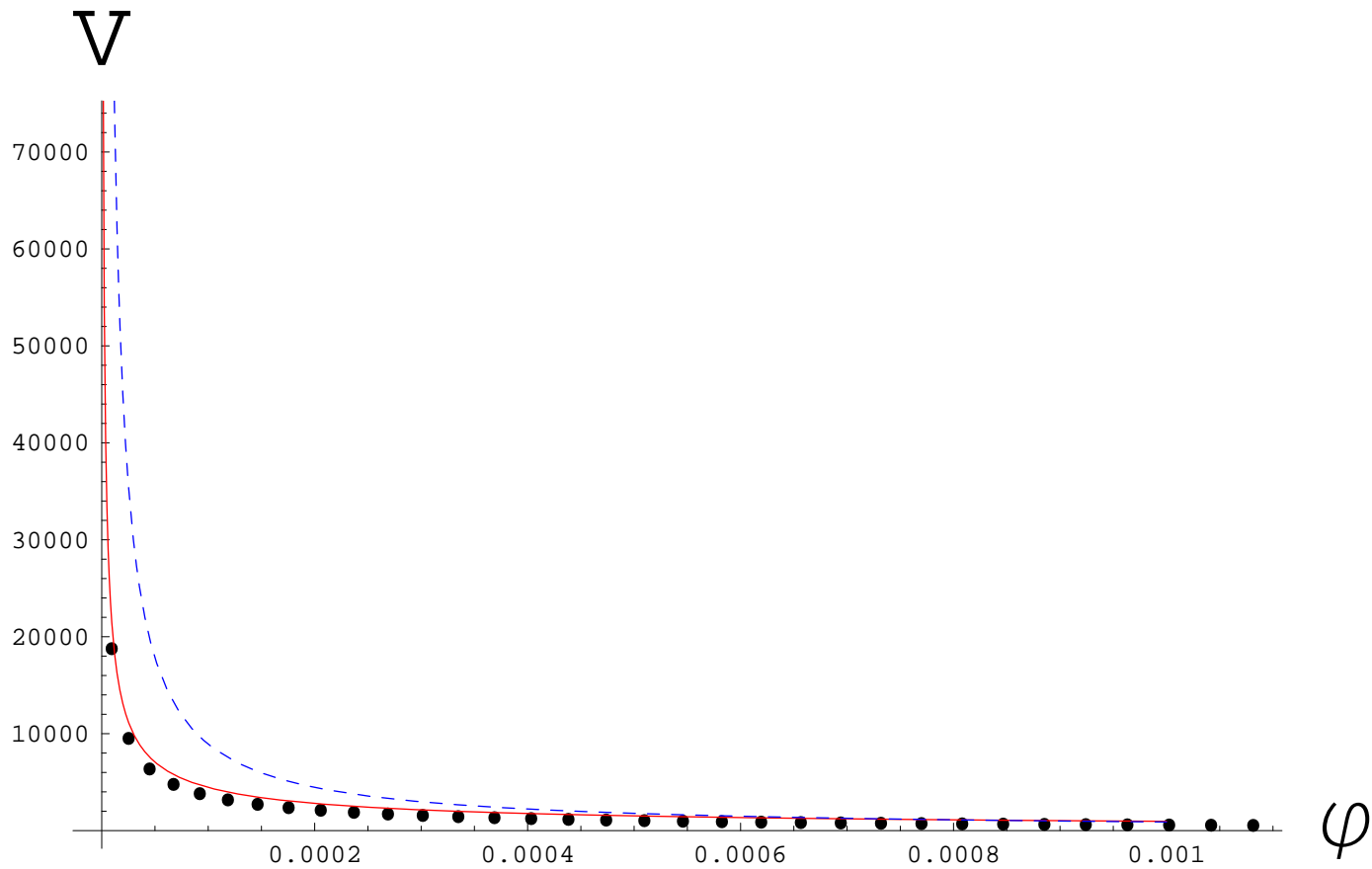
Figure 1: Early-time regime: the dashed curve is the n = 1 scalar field
potential with dust (m = 3); the solid curve is the n = 1 scalar field
potential with radiation (m = 4) and lies closer to the points plotting
the full numerical solution for dust plus radiation and n = 1 scalar field
potential.

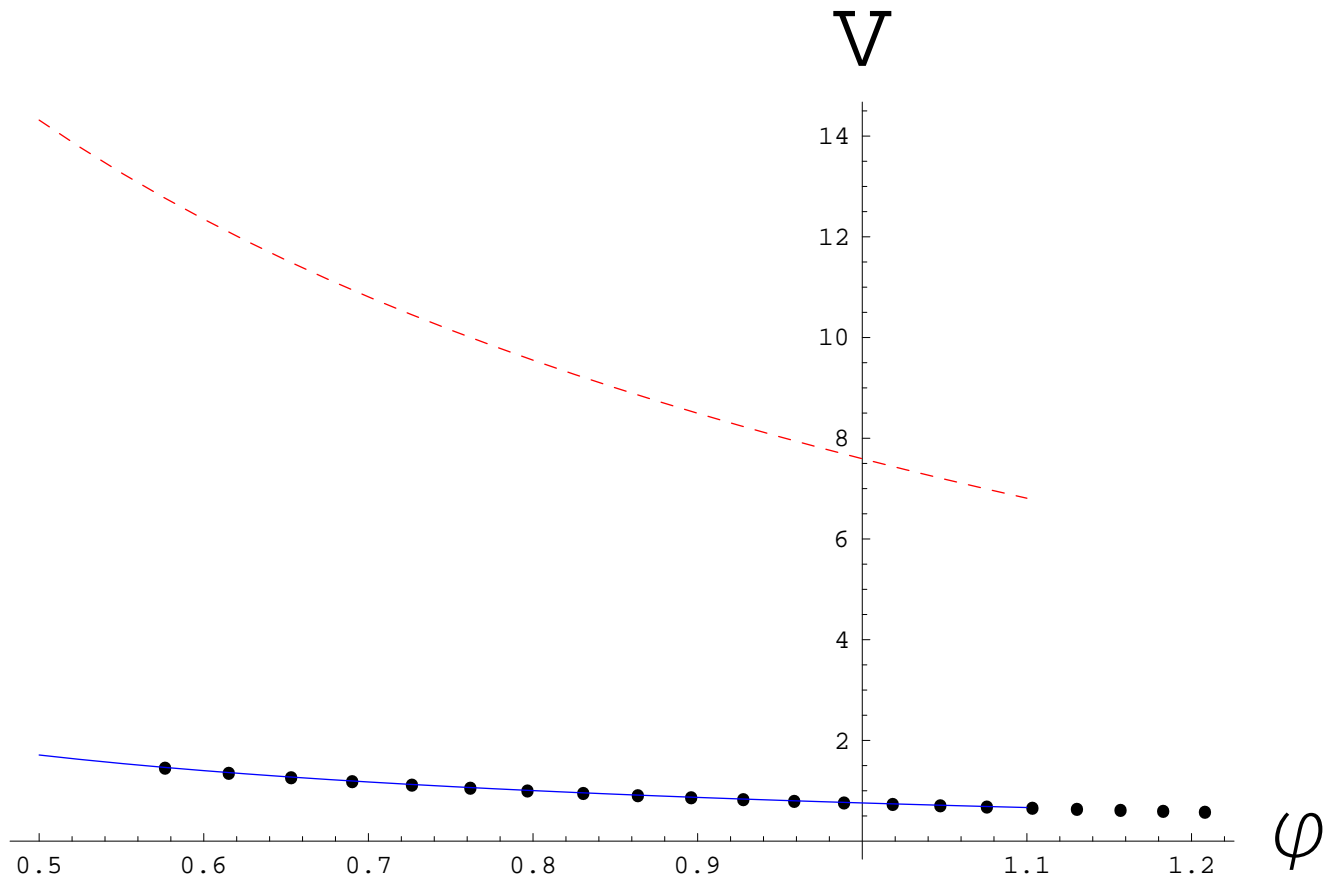
Figure 2: Late-time regime: the dashed curve is the n = 1 scalar field
potential with radiation (m = 4); the solid curve is the n = 1 scalar field
potential with dust (m = 3) and lies closer to the points plotting the full
solution for dust plus radiation and n = 1 scalar field potential.
\end{document}